\documentclass{article}
\usepackage{spconf,amsmath,graphicx}
\usepackage{cite}
\usepackage[colorlinks=true, linkcolor=black, citecolor=black, urlcolor=black]{hyperref}
\usepackage{amssymb}
\usepackage{multirow}
\usepackage{booktabs}
\usepackage{arydshln}

\title{Cross-Lingual F5-TTS: Towards Language-Agnostic Voice Cloning and Speech Synthesis}

\name{%
  \begin{tabular}{c}
    Qingyu Liu$^{1,3}$,
    Yushen Chen$^{1,2}$,
    Zhikang Niu$^{1,2}$,
    Chunhui Wang$^{4}$,
    Yunting Yang$^{4}$,\\
    Bowen Zhang$^{4}$,
    Jian Zhao$^{4}$,
    Pengcheng Zhu$^{4}$,
    Kai Yu$^{1}$,
    Xie Chen$^{1,2,\dagger}$\thanks{$^\dagger$\,Corresponding author}
  \end{tabular}
}

\address{$^1$X-LANCE Lab, MoE Key Lab of Artificial Intelligence, Jiangsu Key Lab of Language Computing, \\Shanghai Jiao Tong University, China    $^2$Shanghai Innovation Institute, China,\\$^3$Johns Hopkins University, USA, $^4$Geely, China}

\begin{document}
\ninept
\maketitle
\begin{abstract}
Flow-matching-based text-to-speech (TTS) models have shown high-quality speech synthesis. However, most current flow-matching-based TTS models still rely on reference transcripts corresponding to the audio prompt for synthesis. This dependency prevents cross-lingual voice cloning when audio prompt transcripts are unavailable, particularly for unseen languages. The key challenges for flow-matching-based TTS models to remove audio prompt transcripts are identifying word boundaries during training and determining appropriate duration during inference. In this paper, we introduce Cross-Lingual F5-TTS, a framework that enables cross-lingual voice cloning without audio prompt transcripts. Our method preprocesses audio prompts by forced alignment to obtain word boundaries, enabling direct synthesis from audio prompts while excluding transcripts during training. To address the duration modeling challenge, we train speaking rate predictors at different linguistic granularities to derive duration from speaker pace. Experiments show that our approach matches the performance of F5-TTS while enabling cross-lingual voice cloning.
\end{abstract}

\section{Introduction}
\label{sec:intro}
Zero-shot Text-to-speech (TTS) aims to generate speech that closely resembles the voice of a given reference speech based on an input text. 
With the scaling of data and model size, existing TTS systems have shown superior capabilities in mimicking speaker characteristics and synthesizing more expressive speech.
The modeling methodologies of zero-shot TTS can be categorized into two types: autoregressive (AR) models\cite{VALLE, SEEDTTS, DiTAR, MiniMaxSpeech, MELLE, CosyVoice2} and non-autoregressive (NAR) models\cite{VOICEBOX, E2TTS, F5TTS, NATURALSPEECH2, NATURALSPEECH3, FASTSPEECH2}.

Among NAR models, flow-matching-based approaches have emerged as particularly promising. For example, VoiceBox\cite{VOICEBOX} surpasses AR-based VALL-E\cite{VALLE} in Word Error Rate (WER) and Speech Similarity (SIM) with 20 times faster inference speed. E2 TTS\cite{E2TTS} and F5-TTS\cite{F5TTS} have achieved remarkable improvements with their simpler architectures. 
Recent cross-lingual TTS approaches, such as VALL-E X~\cite{VALLEX}, have achieved impressive cross-lingual voice cloning capabilities by training on large-scale multilingual data. 
However, all these models still rely on audio prompt transcripts, which creates inherent challenges when reference transcripts are unavailable.
Different flow-matching-based models also employ various duration modeling methods.
While some flow-matching-based models use explicit phoneme duration predictors\cite{VOICEBOX} , F5-TTS estimates duration by the ratio of the audio prompt transcript length and target text length. However, this mechanism breaks down in a cross-lingual context, as text-length ratios between different languages may not correspond to speech duration ratios.

In this paper, we propose Cross-Lingual F5-TTS, a novel framework based on F5-TTS that achieves cross-lingual zero-shot voice cloning by dropping audio prompt transcripts. Leveraging the MMS\cite{MMS} forced alignment, we preprocess our training data to identify word boundaries. 
To solve the challenge of duration prediction, we introduce dedicated speaking rate predictors at three linguistic granularities including phoneme, syllable, and word levels. 
These models are trained to estimate speaking rate from the audio prompt, providing a robust, language-independent mechanism for determining the duration of the target utterance. 
Our approach achieves comparable performance to F5-TTS on LibriSpeech-PC test-clean and Seed-TTS-eval, while successfully extending its capabilities to cross-lingual scenarios with promising results.
\footnote{Audio samples can be found at \url{https://qingyuliu0521.github.io/Cross_lingual-F5-TTS/}. The interactive demo is available at \url{https://huggingface.co/spaces/chenxie95/Cross-Lingual_F5-TTS_Space}.}

\section{Method}
\label{sec:method}
\subsection{Preliminary on Flow-matching-based TTS}
\label{sec:flowmatching}

Flow-Matching-based models\cite{VOICEBOX, E2TTS, F5TTS} have recently achieved remarkable performance in TTS tasks. This approach offers significant advantages in terms of model simplicity. Specifically, by leveraging flow-matching, E2-TTS\cite{E2TTS} and F5-TTS\cite{F5TTS} eliminate additional components such as phoneme duration predictors and complex text encoders, thereby maintaining pipeline simplicity while achieving high-quality synthesis.

The Flow Matching framework aims to learn a time-dependent vector field $v_t$  that matches the probability path $p_t$ between a simple noise distribution $p_0$ and the data distribution $q$ to generate flows $\psi_t$ for sampled flow step $t\in[0,1]$.  The training objective is formulated as the Conditional Flow Matching (CFM) loss:

\vspace{-0.5cm}
\begin{equation}
\mathcal{L}_{\mathrm{CFM}}=\mathbb{E}_{t,q(x_1),p(x_0)}\|v_t(\psi_t(x))-\frac d{dt}\psi_t(x)\|^2
\label{eq:cfm1}
\end{equation}
\vspace{-0.4cm}

This probability path connects a sample $x_0\sim p(x_0)$ from the Gaussian noise to a data sample $x_1\sim q(x_1)$ from the training data.
Optimal Transport (OT) flow matching provides a particularly effective instantiation of this framework. Under the OT formulation, the flow $\psi_t$ is defined as a straight-line trajectory:

\vspace{-0.15cm}
\begin{equation}
\psi_t(x_0) = (1-t)x_0 + tx_1
\label{eq:ot}
\end{equation}
\vspace{-0.3cm}

The corresponding velocity field becomes the constant vector $(x_1 - x_0)$, leading to the OT-CFM loss:

\vspace{-0.4cm}
\begin{equation}
\mathcal{L}_{\mathrm{CFM}}=\mathbb{E}_{t,q(x_1),p(x_0)}\|v_t((1-t)x_0+tx_1)-(x_1-x_0)\|^2
\label{eq:cfm2}
\end{equation}
\vspace{-0.4cm}

In this work, we employ F5-TTS as our baseline. F5-TTS is a fully flow-Matching-based TTS system with diffusion transformer (DiT), which is trained on the text-guided speech-infilling task using OT-CFM. The model predicts masked speech $m \odot x_1$ given its surrounding speech $(1-m)\odot x_1$, the noisy speech $(1-t) x_0+tx_1$ and the extended character sequence $z$.

\subsection{MMS Forced Alignment}
\label{sec:MMS}

To enable transcript-free voice cloning, we leverage the Massively Multilingual Speech (MMS) forced alignment tooling~\cite{MMS} to preprocess our training data and obtain precise word boundary information. The MMS forced alignment system employs a multilingual Wav2Vec2-based acoustic model trained with Connectionist Temporal Classification\cite{CTC} to generate posterior probabilities for audio frames. 
For efficient processing of long audio recordings, MMS chunks audio files into 15-second segments to generate posterior probabilities, then concatenates them back into a unified alignment matrix. 

\begin{figure}[htb]
  \centering
  \includegraphics[width=0.9\linewidth]{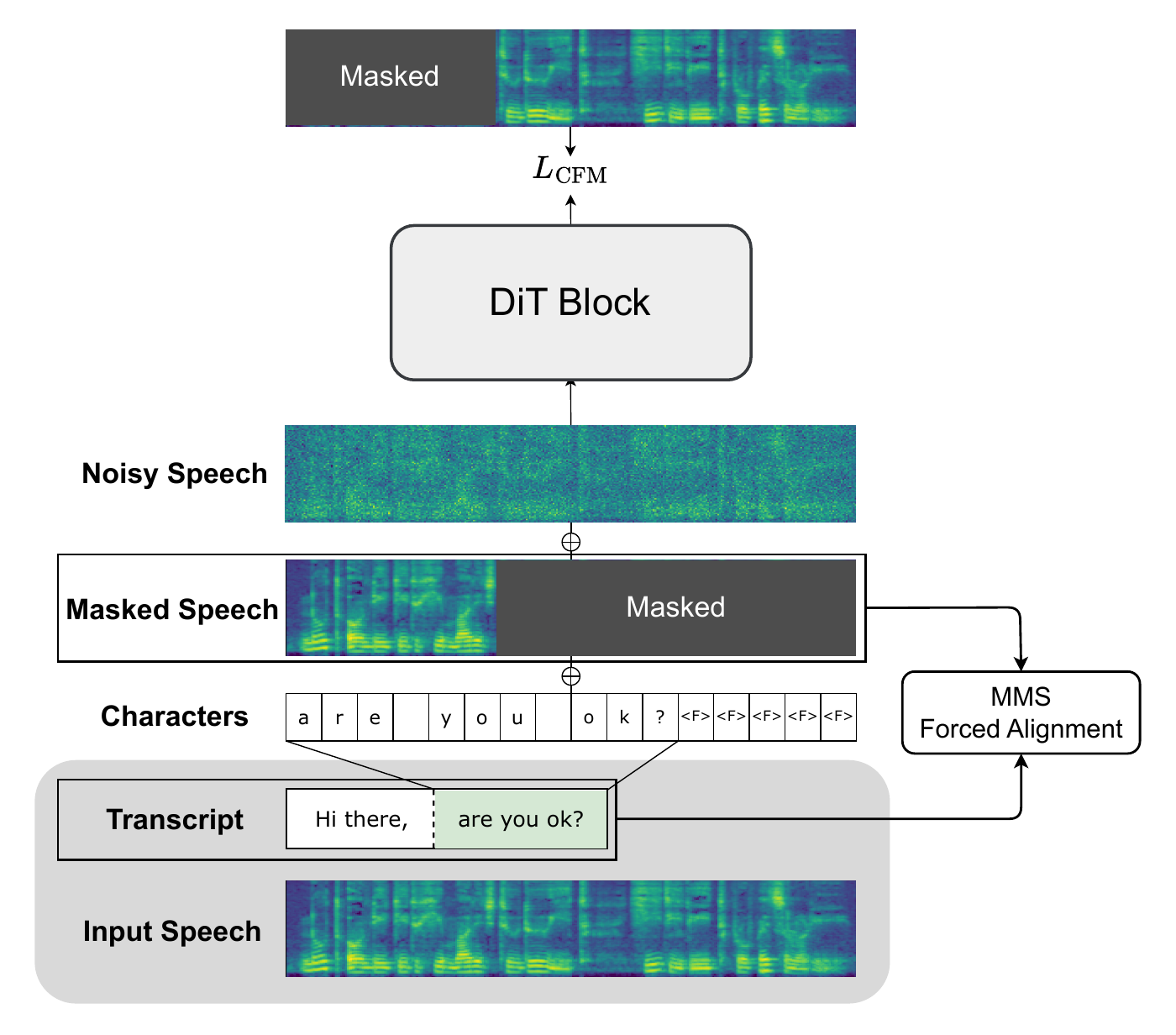}
  \caption{Training framework of Cross-Lingual F5-TTS. MMS forced alignment produces word boundaries for training data, where the left segment serves as transcript-free audio prompt and the mel spectrogram of the right segment is masked for prediction.}
  \label{fig:tts}
\end{figure}

We apply MMS forced alignment to preprocess the Chinese and English subsets of Emilia dataset\cite{Emilia}. For each audio sample and its corresponding transcription, we extract the end time of each word. 
During training, we modify the F5-TTS framework to incorporate this word boundary information as an additional input alongside the original speech and text inputs, as shown in Figure \ref{fig:tts}. At each training step, we randomly select a word boundary within the transcript and partition the audio sample at this point. The audio segment preceding the selected boundary serves as the audio prompt, while we completely discard its corresponding transcript portion. The remaining audio segment is masked and becomes the target for synthesis.

\vspace{-0.1cm}

\subsection{Speaking Rate Predictor}
\label{sec:M}

The elimination of audio prompt transcripts in our transcript-free training approach introduces a critical challenge for duration prediction in speech synthesis.
In the original F5-TTS framework, duration is estimated using a simple length-ratio method where duration equals audio prompt duration multiplied by the ratio of target text length to reference text length. However, this mechanism becomes infeasible when audio prompt transcripts are unavailable. To address this challenge, we introduce a dedicated speaking rate predictor that estimates duration directly from the acoustic characteristics of the audio prompt.

We formulate speaking rate prediction as a discrete classification task and  train three separate models for different linguistic granularities: phonemes per second, syllables per second, and words per second. We define a category set $C$ with uniform intervals of $\Delta = 0.25$. Specifically, the phoneme-level model uses $C = \{0.25, 0.5, \ldots, 17.75, 18.0\}$ with $N = 72$ classes, while both syllable-level and word-level models use $C = \{0.25, 0.5, \ldots, 7.75, 8.0\}$ with $N = 32$ classes. For a given speaking rate $v$, the ground truth category $c_{\text{gt}}$ is determined by minimum distance mapping: $c_{\text{gt}} = \arg\min_{x \in C} |v - x|$. Each model takes audio features as input and independently predicts the corresponding speaking rate category for its respective linguistic unit.

\begin{figure}[htbp]
  \centering
  \includegraphics[width=1.0\linewidth]{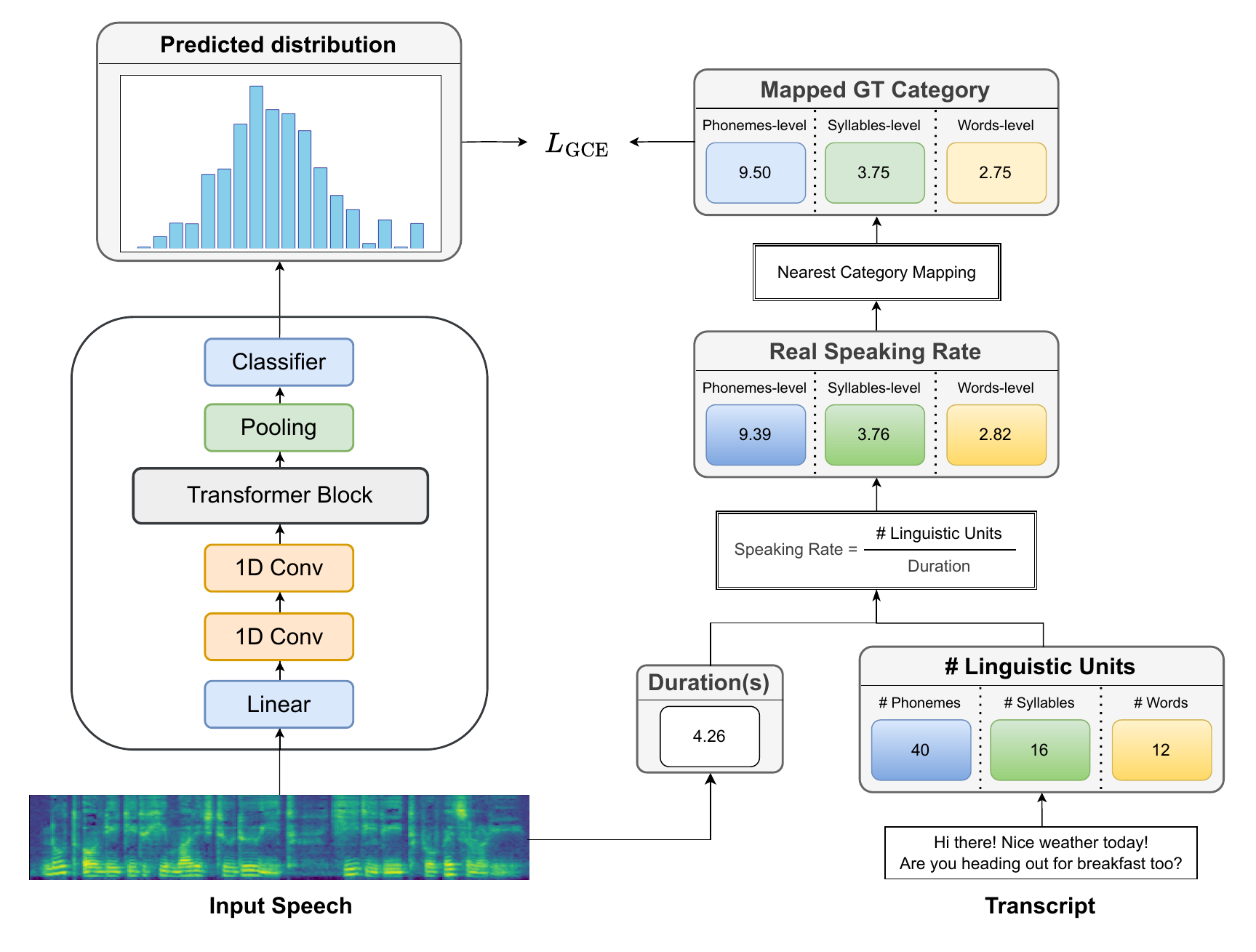}
  \caption{Training pipeline of the speaking rate predictor. The model predicts discrete rate categories from mel-spectrograms, while ground truth speaking rates are mapped to nearest categories for GCE loss computation.}
  \label{fig:sp}
\end{figure}

Our speaking rate predictor employs a transformer-based architecture designed to process mel-spectrogram inputs, as illustrated in Figure \ref{fig:sp}. The model consists of a mel projection layer that projects input mel-spectrograms to the model's hidden dimension, followed by two 1D convolution layers. Multiple transformer encoder layers process the sequence, and an attention-based sequence pooling mechanism aggregates the temporal information by computing attention weights for each time step and performing weighted averaging to obtain a fixed-size representation. Finally, a classifier outputs class probabilities for the speaking rate categories.

Standard cross-entropy loss treats all classes as independent, which is suboptimal for our ordered speaking rate categories. We introduce a Gaussian Cross-Entropy (GCE) loss that incorporates the ordinal nature of speaking rates:

\vspace{-0.3cm}
\begin{equation}
L_{\text{GCE}} = -\frac{1}{N}\sum_{i=1}^{N}\sum_{c=1}^{C} y^{\text{soft}}_{c} \log(\hat{y}_c)
\label{eq:GCE}
\end{equation}
\vspace{-0.2cm}

The soft labels are computed using a Gaussian kernel:

\vspace{-0.15cm}
\begin{equation}
y_c^{\text{soft}}=e^{\frac{-(c-c_{\text{gt}})^2}{2\sigma^2}}
\label{eq:ysoft}
\end{equation}
\vspace{-0.3cm}

where $c_{\text{gt}}$ is the ground truth category index, $c$ is the current category index, and $\sigma$ controls the smoothness of the Gaussian kernel. This formulation assigns higher weights to categories closer to the ground truth, providing tolerance for minor prediction errors while maintaining precision. 

During inference, our approach estimates the target generation duration through a straightforward process. The audio prompt is first fed into the speaking rate predictor to estimate the speaker's characteristic pace in terms of phonemes, syllables, or words per second, and the target text is processed to count the corresponding linguistic units. The target audio duration is then calculated as the ratio of the linguistic unit count to the predicted speaking rate. This mechanism enables our Cross-Lingual F5-TTS to perform duration estimation without relying on audio prompt transcripts, thereby facilitating true cross-lingual voice cloning capabilities.

\section{Experiments}
\label{sec:experiments}

\subsection{Dataset}
\label{sec:dataset}
We train our Cross-Lingual F5-TTS on the Emilia dataset~\cite{Emilia}, a large-scale multilingual speech dataset collected from diverse real-world scenarios. After filtering out transcription failures and misclassified language speech, we retain approximately 95K hours of English and Chinese audio data. To train our speaking rate predictor, we extract a balanced subset containing 500 hours each from the Chinese and English portions of Emilia as training data.

\subsection{Preprocessing}
\label{set:preprocessing}
We apply MMS forced alignment tooling to extract word boundaries for the Emilia dataset. While the Emilia-pipe~\cite{Emilia} employs Whisper-X~\cite{WhisperX} for transcription generation with considerable success, certain challenges remain in the generated transcriptions. Digit numbers, special symbols, and tokens from other languages occasionally appear in the transcriptions, which cannot be properly recognized during the forced alignment process. To address these cases, we implement specialized preprocessing procedures that skip anomalous tokens and exclude them from word boundary extraction, ensuring robust and accurate alignment results.

\subsection{Model Setup}
\label{sec:setup}
Our baseline is open-sourced F5-TTS-Base\cite{F5TTS}, which employs a diffusion transformer (DiT) architecture with 22 layers, 16 attention heads, and 1024 dimensional embeddings. The model is trained for 1.2M updates on eight NVIDIA A100 GPUs with a per-GPU batch size of 38,400 audio frames. We use AdamW optimizer\cite{AdamW} with a learning rate that linearly warms up to $7.5 \times 10^{-5}$ over the first 20k updates, followed by linear decay for the remaining training steps.

Our speaking rate predictor utilizes a transformer-based architecture with 6 layers, 8 attention heads, and 512 dimensional embeddings. Training is conducted on four A100 GPUs for 50k updates with a per-GPU batch size of 38,400 audio frames. The learning rate is warmed up to $2.5 \times 10^{-4}$ over the first 7.5k updates and then linearly decayed. For the Gaussian Cross-Entropy loss, we set the standard deviation $\sigma = 1.0$ to balance between tolerance for nearby predictions and precision for the target category.

For inference, we follow the standard F5-TTS settings using Euler ODE solver with 32 function evaluations (NFE = 32), CFG strength 2.0, sway sampling coefficient -1.0, and pre-trained Vocos\cite{VOCOS} as the vocoder.

\subsection{Evaluation}
\label{sec:evaluation}

We follow the evaluation setting of F5-TTS, adopting Seed-TTS-eval and LibriSpeech-PC test-clean\cite{LIBRISPEECHPC} as our test set. We also build a multilingual cross-lingual test set with 473 samples of 3-8 second audio prompts from FLEURS\cite{FLEURS}, covering four languages (German, French, Hindi, Korean) to synthesize both English and Chinese speech. The evaluation is conducted using the following three metrics:

\noindent\textbf{Word Error Rate (WER)} measures the intelligibility of synthesized speech by comparing its transcription with the ground truth text. We employ Whisper-large-V3\cite{WHISPER} and Paraformer-zh\cite{PARAFORMER} for automatic recognition and compute WER accordingly.

\noindent\textbf{Speaker Similarity (SIM-o)} quantifies the resemblance between the synthesized and the original target speeches. We use WavLM-large-based\cite{WAVLM} speaker verification model to extract speaker embeddings and compute the cosine similarity between them.

\noindent\textbf{UTMOS}\cite{UTMOS} provides an automated assessment of speech naturalness through a pre-trained MOS prediction model. It estimates audio quality without requiring reference recordings or labels. While not an absolute subjective measure, UTMOS provides a practical and efficient way to evaluate the naturalness of synthetic speech.


To evaluate the effectiveness of our speaking rate predictor, we assess its duration prediction accuracy using two complementary metrics:

\noindent\textbf{Mean Relative Error (MRE)} measures the relative accuracy of duration prediction. It is calculated as the average relative difference between the predicted duration (derived from the number of linguistic units divided by the predicted speaking rate) and the ground truth audio duration.

\noindent\textbf{Mean Absolute Error (MAE)} quantifies the absolute deviation between the predicted and ground truth duration.

\section{Results}
\label{sec:results}

\subsection{Speaking Rate Predictor Performance}
\label{sec:M_results}
Table~\ref{tab:mae_mre_results} presents the duration prediction accuracy results of our speaking rate predictors across different linguistic granularities.
M1 (phoneme-level predictor) outperforms other predictors on LibriSpeech-PC test-clean and shows comparable results to M2 on SeedTTS test-en, demonstrating its effectiveness for English duration modeling. Notably, M2 (syllable-level predictor) performs optimally on SeedTTS test-zh, suggesting that syllables may be more natural linguistic units for duration modeling in Chinese. 
The consistently inferior performance of M3 (word-level predictor) across all datasets highlights the limitations of coarse-grained linguistic units. 

\begin{table}[!htbp]
\renewcommand\arraystretch{1.2}
\caption{MAE (s) and MRE (\%) of predicted durations for different speaking rate predictors.}
\label{tab:mae_mre_results}
\vspace{+0.2cm}
\centering
\footnotesize
\begin{tabular}{llcc}
\hline
\textbf{ID} & \textbf{System} & \textbf{MAE(s)$\downarrow$} & \textbf{MRE(\%)$\downarrow$}\\
\hline
\multicolumn{4}{c}{\textbf{Librispeech-PC \textit{test-clean}}} \\
\hline
\textbf{M1} & Phonemes-level predictor & 0.759 & \textbf{11.932} \\
\textbf{M2} & Syllables-level predictor & \textbf{0.757} & 11.945 \\
\textbf{M3} & Words-level predictor & 1.171 & 18.406 \\
\hline
\multicolumn{4}{c}{\textbf{SeedTTS \textit{test-en}}} \\
\hline
\textbf{M1} & Phonemes-level predictor & \textbf{0.637} & \textbf{15.017} \\
\textbf{M2} & Syllables-level predictor & 0.704 & 16.497 \\
\textbf{M3} & Words-level predictor & 0.886 & 20.031 \\
\hline
\multicolumn{4}{c}{\textbf{SeedTTS \textit{test-zh}}} \\
\hline
\textbf{M1} & Phonemes-level predictor & 0.845 & 14.469 \\
\textbf{M2} & Syllables-level predictor & \textbf{0.783} & \textbf{13.771} \\
\textbf{M3} & Words-level predictor & 0.908 & 16.156 \\
\hline
\end{tabular}
\vspace{-0.3cm}
\end{table}

\subsection{Intra-lingual Voice Cloning Performance}
\label{sec:intra_results}
Table~\ref{tab:all_results} compares our Cross-Lingual F5-TTS (CL-F5) with the original F5-TTS baseline on standard benchmarks. Our method demonstrates competitive performance across all evaluation metrics. On LibriSpeech-PC test-clean, CL-F5 with M1 and M2 achieves superior WER and UTMOS compared to the baseline. This trend continues on SeedTTS test-en, where CL-F5 with M1 delivers better WER and UTMOS than the baseline. Although SIM is slightly lower for CL-F5 compared to the baseline on the English test sets, this modest decrease is an acceptable trade-off for the extension of cross-lingual capability. On SeedTTS test-zh, CL-F5 with M2 demonstrates performance almost identical to the baseline across all metrics. 
The language-specific preferences observed in Section~\ref{sec:M_results} translate to the intra-lingual voice cloning: CL-F5 with fine-grained predictors (M1, M2) consistently outperforms CL-F5 with M3. 
These results fully validate the effectiveness of our method that CL-F5 retains intra-lingual voice cloning quality while eliminating the dependency on audio prompt transcripts.

\vspace{-0.1cm}
\begin{table}[!htbp]
\renewcommand\arraystretch{1.2}
\caption{Results on LibriSpeech-PC test-clean, SeedTTS test-en, and SeedTTS test-zh.}
\label{tab:all_results}
\vspace{+0.2cm}
\centering
\footnotesize
\begin{tabular}{llccc}
\hline
\textbf{System} & \textbf{Duration Method} & \textbf{WER(\%)$\downarrow$} & \textbf{SIM-o$\uparrow$} & \textbf{UTMOS$\uparrow$}\\
\hline
\multicolumn{5}{c}{\textbf{LibriSpeech-PC \textit{test-clean}}} \\
\hline
Baseline & Length-ratio & 2.205 & \textbf{0.668} & 3.797 \\
\hdashline
\textbf{CL-F5} & \textbf{M1} & \textbf{2.079} & 0.663 & 3.884 \\
\textbf{CL-F5} & \textbf{M2} & 2.120 & 0.658 & \textbf{3.892} \\
\textbf{CL-F5} & \textbf{M3} & 2.894 & 0.652 & 3.855 \\
\hline
\multicolumn{5}{c}{\textbf{SeedTTS \textit{test-en}}} \\
\hline
Baseline & Length-ratio & 1.545 & \textbf{0.676} & 3.581 \\
\hdashline
\textbf{CL-F5} & \textbf{M1} & \textbf{1.513} & 0.662 & \textbf{3.629} \\
\textbf{CL-F5} & \textbf{M2} & 1.594 & 0.660 & 3.625 \\
\textbf{CL-F5} & \textbf{M3} & 2.009 & 0.646 & 3.593 \\
\hline
\multicolumn{5}{c}{\textbf{SeedTTS \textit{test-zh}}} \\
\hline
Baseline & Length-ratio & \textbf{1.475} & 0.762 & 2.898 \\
\hdashline
\textbf{CL-F5} & \textbf{M1} & 1.605 & 0.759 & \textbf{2.913} \\
\textbf{CL-F5} & \textbf{M2} & 1.481 & \textbf{0.764} & 2.887 \\
\textbf{CL-F5} & \textbf{M3} & 1.616 & 0.763 & 2.889 \\
\hline
\end{tabular}
\vspace{-0.3cm}
\end{table}

\subsection{Cross-lingual Voice Cloning Results}
\label{sec:cross_results}
Table~\ref{tab:cross} presents the cross-lingual voice cloning results on our multilingual test set, where the audio prompt and target synthesis languages differ. 
The cross-lingual results reinforce the language-specific preferences observed in Section~\ref{sec:M_results} and Section~\ref{sec:intra_results}: CL-F5 achieves optimal performance with M1 for English targets and M2 for Chinese targets. CL-F5 with M3 shows significant performance degradation in cross-lingual scenarios, with WER dramatically increasing to 16.494\% for English targets compared to 2.496\% for CL-F5 with M1. This is because M3 produces overly fast speaking rate, leading to compressed temporal patterns that negatively impact intelligibility. These results highlight the critical importance of accurate duration modeling in cross-lingual TTS, as temporal inaccuracies can severely degrade overall synthesis quality.

Remarkably, our approach demonstrates effective cross-lingual voice cloning capabilities despite the speaking rate predictors being trained exclusively on Chinese and English data, showing good generalization to unseen languages without requiring audio prompt transcripts.

\begin{table}[!htbp]
\renewcommand\arraystretch{1.2}
\caption{Cross-lingual voice cloning results of Cross-Lingual F5-TTS. GT length denotes the result obtained by using ground truth total speech length. }
\label{tab:cross}
\vspace{+0.2cm}
\centering
\footnotesize
\begin{tabular}{lccc}
\hline
\textbf{Duration Method} & \textbf{WER(\%)$\downarrow$} & \textbf{SIM-o$\uparrow$} & \textbf{UTMOS$\uparrow$}\\
\hline
\multicolumn{4}{c}{\textbf{Cross-lingual \textit{test-en}}} \\
\hline
GT length & \textbf{2.462} & 0.530 & \textbf{3.083} \\
\hdashline
\textbf{M1} & 2.496 & \textbf{0.543} & 3.069 \\
\textbf{M2} & 4.362 & 0.518 & 3.059 \\
\textbf{M3} & 16.494 & 0.486 & 2.926 \\
\hline
\multicolumn{4}{c}{\textbf{Cross-lingual \textit{test-zh}}} \\
\hline
GT length & \textbf{1.596} & 0.558 & 2.452 \\
\hdashline
\textbf{M1} & 2.446 & 0.555 & 2.494 \\
\textbf{M2} & 1.801 & \textbf{0.565} & \textbf{2.503} \\
\textbf{M3} & 1.946 & 0.563 & 2.492 \\
\hline
\end{tabular}
\vspace{-0.3cm}
\end{table}

\section{Conclusion}
\label{sec:conclusion}
In this work, we present Cross-Lingual F5-TTS, a novel framework that achieves cross-lingual zero-shot voice cloning by eliminating the dependency on audio prompt transcripts. Our approach addresses a fundamental limitation of existing NAR TTS models by leveraging MMS forced alignment to partition training data at word boundaries. To solve the consequent duration prediction challenge, we introduce dedicated speaking rate predictors that estimate duration directly from acoustic features using phoneme, syllable, or word-level granularities. Experimental results demonstrate that our method maintains comparable performance to the original F5-TTS on LibriSpeech-PC test-clean and Seed-TTS-eval, while successfully enabling cross-lingual voice cloning from speakers of unseen languages that the baseline cannot handle. The speaking rate predictor analysis reveals that finer-grained linguistic units including phoneme and syllable provide more reliable duration estimation. 
Although our transcript-free approach enables cross-lingual voice cloning, it shows reduced capability in transferring subtle speaker characteristics such as accents and emotions compared to the original F5-TTS. In future work, we plan to explore methods to compensate for the missing linguistic information from audio prompt transcripts, enabling Cross-Lingual F5-TTS to generate more expressive speech while maintaining its cross-lingual capabilities.

\section{Acknowledgments}
This work was supported by the Science and Technology Innovation (STI) 2030-Major Project (2022ZD0208700), National Natural Science Foundation of China  (No. U23B2018), Shanghai Municipal Science and Technology Major Project under Grant 2021SHZDZX0102 and Yangtze River Delta Science and Technology Innovation Community Joint Research Project (2024CSJGG110 0).

The authors acknowledge Beijing PARATERA Tech CO., Ltd. for providing HPC resources that have contributed to the research results reported within this paper. URL: \url{https://paratera.com/}


\bibliographystyle{IEEEbib}
\let\oldbibliography\thebibliography
\renewcommand{\thebibliography}[1]{%
  \oldbibliography{#1}%
  \renewcommand{\baselinestretch}{0.9}%
  \small%
  \setlength{\itemsep}{0pt}%
  \setlength{\parskip}{0pt}%
}
\bibliography{refs}

\end{document}